\documentclass[twocolumn,superscriptaddress,amsmath,amssymb,aps,prl]{revtex4-1}
\usepackage{graphicx}
\usepackage{amsmath}

\begin{document}

\title{Realizing the Hayden-Preskill Protocol with Coupled Dicke Models} 

\author{Yanting Cheng}
\thanks{They contribute equally to this work. }
\affiliation{Institute for Advanced Study, Tsinghua University, Beijing, 100084, China}

\author{Chang Liu}
\thanks{They contribute equally to this work. }
\affiliation{Institute for Advanced Study, Tsinghua University, Beijing, 100084, China}

\author{Jinkang Guo}
\thanks{They contribute equally to this work. }
\affiliation{Institute for Advanced Study, Tsinghua University, Beijing, 100084, China}
\affiliation{Department of Physics, Peking University, Beijing 100871, China}

\author{Yu Chen}
\affiliation{Center for Theoretical Physics, Department of Physics, Capital Normal University, Beijing, 100048, China}

\author{Pengfei Zhang}
\affiliation{Institute for Advanced Study, Tsinghua University, Beijing, 100084, China}
\affiliation{Institute for Quantum Information and Matter, California Institute of Technology, Pasadena, California 91125, USA}
\affiliation{Walter Burke Institute for Theoretical Physics, California Institute of Technology, Pasadena, California 91125, USA}

\author{Hui Zhai}
\email{hzhai@tsinghua.edu.cn}
\affiliation{Institute for Advanced Study, Tsinghua University, Beijing, 100084, China}

\date{\today}

\begin{abstract}
Hayden and Preskill proposed a thought experiment that Bob can recover the information Alice throws into a black hole if he has a quantum computer entangled with the black hole, and Yoshida and Kitaev recently proposed a concrete decoding scheme. In the context of quantum many-body physics, the parallel question is that after a small system is thermalized with a large system, how one can decode the initial state information with the help of two entangled many-body systems. Here we propose to realize this decoding protocol in a physical system of two Dicke models, with two cavity fields prepared in a thermofield double state. We show that the Yoshida-Kitaev protocol allows us to read out the initial spin information after it is scrambled into the cavity. We show that the readout efficiency reaches a maximum when the model parameter is tuned to the regime where the system is the most chaotic,  characterized by the shortest scrambling time in the out-of-time-ordered correlation function. Our proposal opens up the possibility of discussing this profound thought experiment in a realistic setting. 
\end{abstract}
\maketitle

Quantum information scrambling now plays an important role in understanding the quantum many-body system. When a many-body system evolving from an initial state finally thermalizes, all the local information about the initial state gets lost, since a thermalized many-body system is described by only a few parameters such as temperature and chemical potential \cite{ETH1,ETH2,ETH3,ETH4}. Precisely speaking, the local information of the initial state has been scrambled into the entire system during the process of quantum thermalization, such that the retrieval of this local information from local measurements is not possible \cite{Page,HP}. This information loss is reminiscent of the black hole information problem. When Alice throws her diary into the black hole, Bob can not recover the information in the diary from the Hawking radiation, which is just a small portion of the entire Hilbert space of a black hole. Here the black hole is considered as the fastest information scrambler in our universe \cite{bound,Blackhole}. Recently, the out-of-time-ordered correlation (OTOC) function has been studied for describing the information scrambling processes, and a Lyapunov exponent can be defined to characterize the speed of information scrambling \cite{Blackhole,Lukin,Kitaev1,XL,OR}. It is now known that a black hole possesses the largest Lyapunov exponent \cite{bound,Blackhole}.

In a seminal paper, Hayden and Preskill proposed a thought experiment for Bob to recover the information that Alice threw into a black hole, which is now known as the Hayden-Preskill (HP) protocol \cite{HP}. The key of the HP protocol is to have another quantum system maximally entangled with the black hole and entirely under Bob's control. Based on general quantum information theory, Hayden and Preskill show that in this setting, it is possible for Bob to recover the information in Alice's diary by only manipulating the number of qubits much less than the total system \cite{HP}. In the parallel discussion of quantum many-body system, it means that the initial state information can be recovered even after thermalization if two entangled many-body systems are prepared.  

\begin{figure}[t]
\centering
\includegraphics[width=0.95\linewidth]{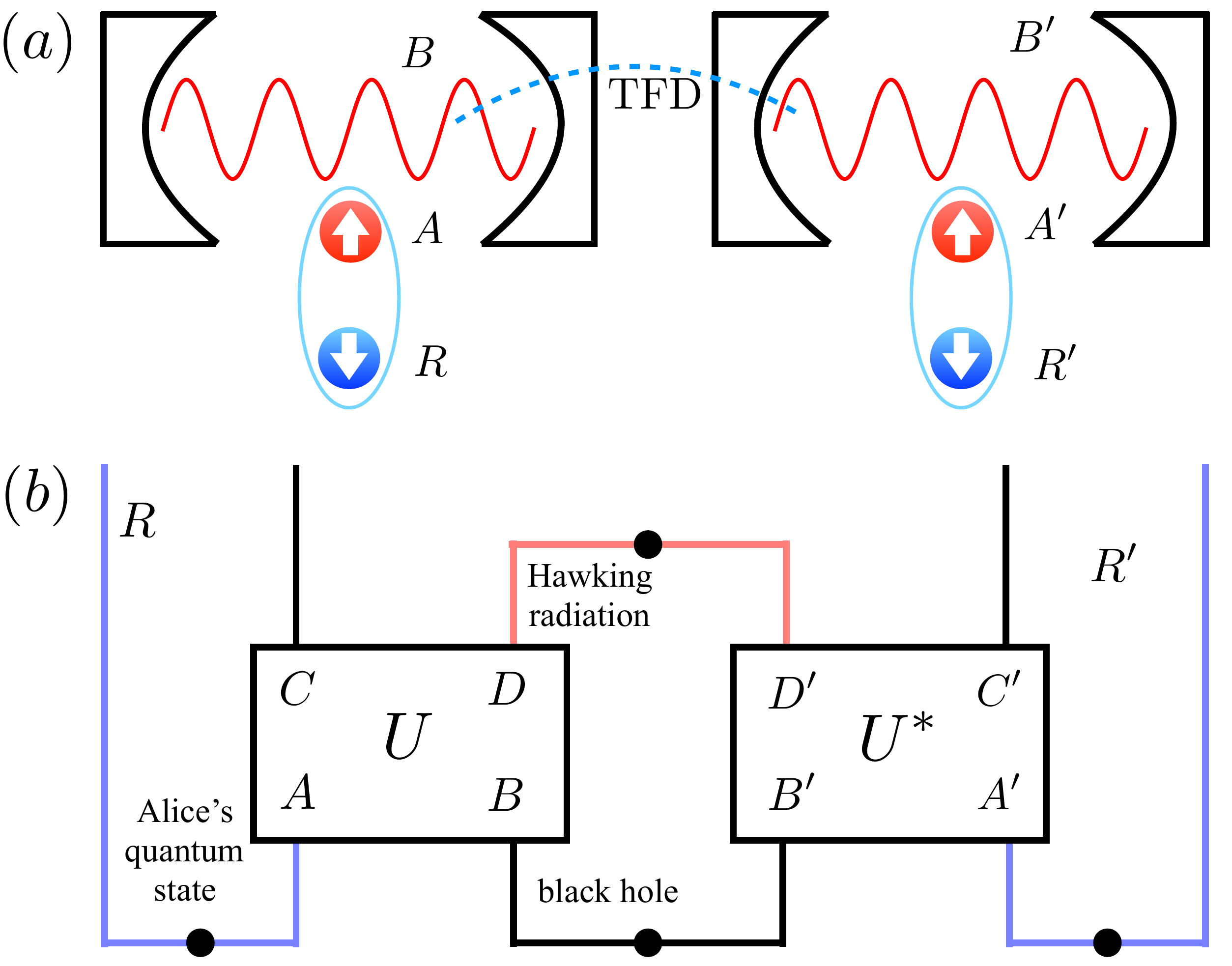}
\caption{(a) Schematic of our physical system of coupled Dicke model, where ``TFD" stands for the thermofield double state.  (b) Diagrammatical illustration of the Yoshida and Kitaev version of the Hayden-Preskill protocol. }
\label{system}
\end{figure}
	
Recently, Yoshida and Kitaev (YK) put a step forward and describe a procedure of how to realize the HP protocol \cite{YK}. Their protocol requires the evolution of the quantum system described by a Haar-random unitary evolution \cite{fast}, which perfectly scrambles the information as a black hole. They also require that the unitary evolutions of two entangled many-body systems are conjugate of each other \cite{YK}. In this letter, we propose a concrete physical model for realizing the YK version of HP protocol (short-noted as YKHP protocol hereafter) using two coupled Dicke models, as shown in Fig. \ref{system}(a). The quantum evolution of a physical system is governed by its Hamiltonian. Although thermalization and information scrambling can happen in most Hamiltonian systems, except for a few exceptions like many-body localization system, most systems are not as chaotic as a Haar-random, and the information scrambling in these systems is not as fast as a black hole. On the other hand, for the two systems evolving under two conjugating unitary evolutions, the Hamiltonians for these two systems have to be opposite with each other. Admittedly, in many synthetic quantum systems, parameters in the Hamiltonian are largely tunable, we should also anticipant that two Hamiltonians can not be perfectly opposite with each other. These are major difference between the physical system and the ideal situation considered by YK. The purpose of this work is to investigate how all these practical effects influence the efficiency of the YKHP protocol.

\textit{YKHP Protocol.} Before starting the discussion of our physical realization, let us first briefly review the YKHP protocol \cite{HP,YK}, which is diagrammatically illustrated in Fig. \ref{system}(b). Here ``$\mathcal{A}$" stands for a small quantum state with Hilbert space dimension $d_\mathcal{A}$ hold by Alice, and ``$\mathcal{B}$" stands for a black hole as a large quantum system with Hilbert space dimension $d_\mathcal{B}$. ``$\mathcal{D}$" stands for Hawking radiation as a small part of the black hole with Hilbert space $d_\mathcal{D}$. Their Hilbert space dimensions satisfy the condition that $d_\mathcal{A}\ll d_\mathcal{D}\ll d_\mathcal{B}$.  Here throwing Alice's diary into the black hole means that the small system $\mathcal{A}$ is coupled to a large system $\mathcal{B}$, after which the total system undergoes a Haar-random unitary evolution denoted by $\mathcal{U}$. 

Before Alice couples her system $\mathcal{A}$ with the large system $\mathcal{B}$, she first backed up her information by forming an EPR pair between $\mathcal{A}$ and another reference system denoted by ``$\mathcal{R}$". Systems $\mathcal{R}$ and $\mathcal{A}$ have the same Hilbert space dimensions. Because system-$\mathcal{R}$ is never coupled to the large system $\mathcal{B}$ and always remains independent during the evolution, and system-$\mathcal{A}$ is coupled to $\mathcal{B}$, the EPR correlation between $\mathcal{A}$ and $\mathcal{R}$ smears out after thermalization. Thus, the so-called retrieval Alice's information is formulated as whether one can recover the EPR correlation between sub-system $\mathcal{D}$ and $\mathcal{D'}$ and the reference system $\mathcal{R}$ and $\mathcal{R'}$. If without the right-half systems in Fig. \ref{system}(b) and Bob is only allowed to perform measurement only on sub-system $\mathcal{D}$, it is not possible to recover Alice's information because it can be shown that the mutual information between $\mathcal{D}$ and $\mathcal{R}$ is strictly zero after information is completely scrambled. 

In this protocol, we need to introduce another half system denoted by $\mathcal{A}^\prime$, $\mathcal{B}^\prime$, $\mathcal{D}^\prime$, and $\mathcal{R}^\prime$, which have the same Hilbert space dimensions as $\mathcal{A}$, $\mathcal{B}$, $\mathcal{D}$, and $\mathcal{R}$, respectively. Initially, $\mathcal{B}$ and $\mathcal{B}^\prime$ form a maximally entangled state, and $\mathcal{A}^\prime$ and $\mathcal{R}^\prime$ form the same EPR correlation as that between $\mathcal{A}$ and $\mathcal{R}$. When $\mathcal{A}$ is coupled to $\mathcal{B}$, simultaneously $\mathcal{A}^\prime$ is also coupled to $\mathcal{B}^\prime$. The total system including $\mathcal{A}^\prime$ and $\mathcal{B}^\prime$ undergoes unitary evolution $\mathcal{U}^*$ conjugating with $\mathcal{U}$. The essential point of YK's paper is that when $\mathcal{D}$ and $\mathcal{D}^\prime$ is projected to an EPR state, the initial EPR correlation will be recovered between $\mathcal{R}$ and $\mathcal{R}^\prime$. That is to say, let us use $\mathcal{F}$ to denote the conditional probability for $\mathcal{R}$ and $\mathcal{R}^\prime$ being in the EPR state when $\mathcal{D}$ and $\mathcal{D}^\prime$ is projected to an EPR state, they show $\mathcal{F}=1$ when $\mathcal{U}$ and $\mathcal{U}^*$ are Haar-random unitary and fully scramble the information. 

\begin{figure}[t]
\centering
\includegraphics[width=0.95\linewidth]{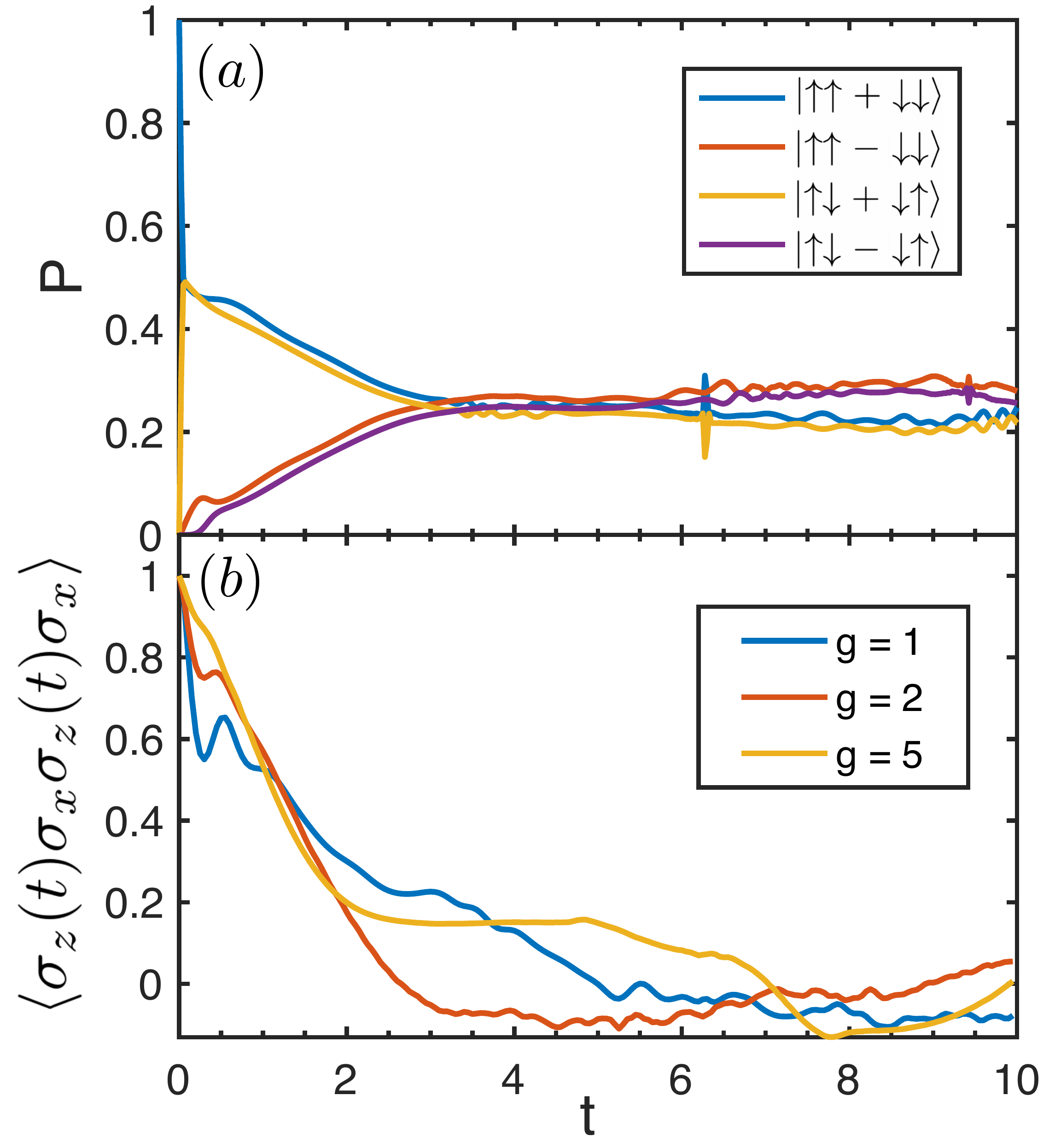}
\caption{Thermalization and OTOC for a single Dicke model. (a) The probability of projection onto the four Bell states formed by $\mathcal{A}$ and $\mathcal{R}$. (b) OTOC for the Dicke model with three different coupling constant $g/\hbar\omega_0$. Both (a) and (b) are plotted in term of $t$ (in unit of $1/\omega_0$). Here $\beta=0.01/\hbar\omega_0$ and $\hbar\omega_z=3\hbar\omega_0$.  }
\label{thermalization}
\end{figure}

\textit{Coupled Dicke Model.} Dicke model describes a spin coupled to a single mode cavity photon field. The Hamiltonian for a single Dicke model is written as
\begin{equation}
\hat{H}_\text{Dicke}(\hat{a},\sigma)=\hbar\omega_0\hat{a}^\dag\hat{a}+g(\hat{a}^\dag+\hat{a})\sigma_x+\hbar\omega_z\sigma_z. 
\end{equation}
where $\sigma$ is the Pauli matrices for a spin-$1/2$ and $\hat{a}$ is the cavity photon field, and we use $\hbar\omega_0$ as unit of energy. We choose the Dicke model to realize the HP protocol for following reasons. 

Firstly, the spin-$1/2$ plays the role as Alice's small system $\mathcal{A}$.  
The Hilbert space for the cavity field is the Fock space spanned by $\{|n\rangle, n=0,\dots,\infty\}$, whose dimension is much larger than that of the Hilbert space $\mathcal{A}$, and the cavity field plays the role as the black hole $\mathcal{B}$. Suppose initially $g=0$ and the spin and the cavity are decoupled, then we turn on the coupling $g$ at $t=0$ and this quench process naturally mimics the processes that Alice's diary is thrown into the black hole. The quantum evolution under $\hat{H}_\text{Dicke}$ simulates $\mathcal{U}$. 

Secondly, in $\hat{H}_\text{Dicke}$, $\hbar\omega_0$ is the cavity photon energy detuning comparing with the background pumping laser field \cite{squeeze1}, $g$ is the cavity-atom coupling and $\hbar\omega_z$ is the Zeeman field for spin. All these parameters can be tuned over wide ranges and their signs can be changed. By inverting the signs of all three parameters, one can realize $-\hat{H}_\text{Dicke}$, under which the evolution realizes $\mathcal{U}^*$. 

Thirdly, one can generate a thermofield double (TFD) state between two cavity photons by a pair creation process, or a two-mode squeezing process known in quantum optics \cite{squeeze1,squeeze2,Chin}. Here we consider two Dicke models with $\hat{H}=\hat{H}_\text{L}+\hat{H}_\text{R}$, where $\hat{H}_\text{L}=\hat{H}_\text{Dicke}(\hat{a}_\text{L},\sigma_\text{L})$ and $\hat{H}_\text{R}=-\hat{H}_\text{Dicke}(\hat{a}_\text{R},\sigma_\text{R})$, where $\hat{a}_\text{L}$ and $\hat{a}_\text{R}$ are photon field operators for the left and the right cavity fields, and $\sigma_\text{L}$ and $\sigma_\text{R}$ are two spin-$1/2$ for $\mathcal{A}$ and $\mathcal{A}^\prime$. The Hamiltonian does not couple two sides, and the entanglement between two Dicke models is established through the initial state, which is prepared in a TFD state defined as \cite{TFD1,TFD2}
\begin{equation}
|\text{TFD}_{\mathcal{B}\mathcal{B}^\prime}\rangle=\frac{1}{\sqrt{Z}}\sum\limits_{n}e^{-n\beta\hbar\omega_0/2}|n\rangle_\text{L}|n\rangle_\text{R},
\end{equation}   
where $Z=\sum_n e^{-\beta n\hbar\omega_0}$. The $\beta\rightarrow 0$ limit of the TFD state is the maximally entangled state used in YK's paper. Here we always focus on the finite $\beta$ case, because the Hilbert space dimension for cavity field is unbounded, and finite $\beta$ naturally introduces a cut-off such that our numerical calculation below can be safely done in a finite Hilbert space. It can be shown that YK's proposal also works for a TFD state at small but finite $\beta$ \cite{supple}.   
 
\textit{Thermalization and OTOC.} We first consider a single Dicke model. Initially, we set $g=0$ where the spin and cavity is decoupled. We set that the cavity is in a thermal equilibrium state with $\beta=0.01$, and $\mathcal{A}$ and $\mathcal{R}$ are in a EPR state $|\Psi^\text{EPR}_{\mathcal{A}\mathcal{R}}\rangle=\frac{1}{\sqrt{2}}\left(|\uparrow\rangle_\mathcal{A}|\uparrow\rangle_\mathcal{R}+|\downarrow\rangle_\mathcal{A}|\downarrow\rangle_\mathcal{R}\right)$, which is one of the four Bell states. Then at $t=0$ we turn on finite coupling $g$ and let the system evolve. For $\beta=0.01$ we considered here, we verified that it is safe to truncate the Hilbert space to several hundreds, and with this truncation we can numerically calculate the full quantum dynamics and obtain all correlation functions. 

\begin{figure}[t]
\centering
\includegraphics[width=0.95\linewidth]{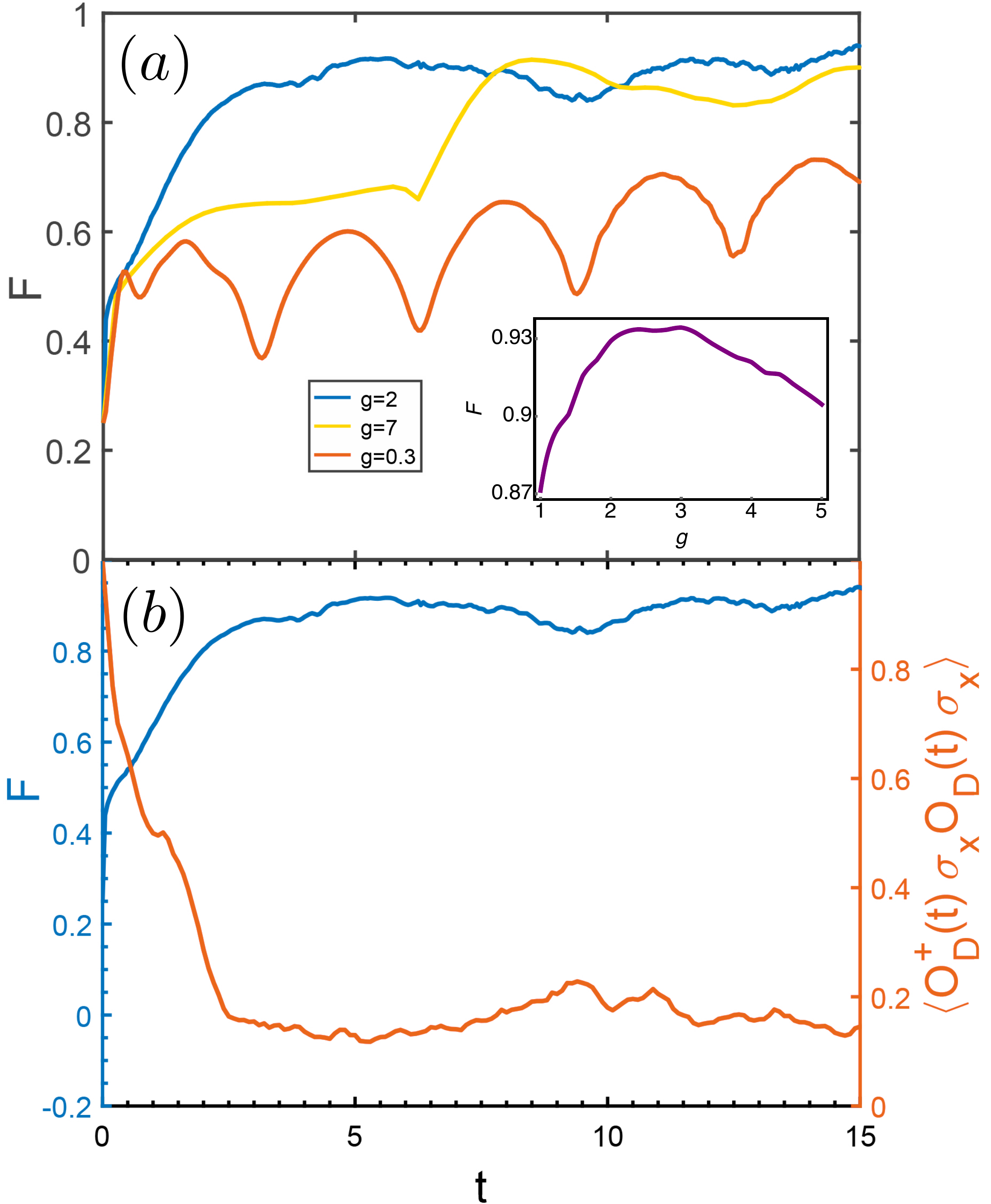}
\caption{(a) The conditional probability $\mathcal{F}$ as a function of $t$ for three different coupling constant $g$. (b) $\mathcal{F}$ and an OTOC between spin and an operator $\hat{\mathcal{O}}_\mathcal{D}$ in Hilbert space $\mathcal{D}$ are plotted in the same figure with the same coupling constant $g=2$.  }
\label{conditional_probability}
\end{figure}

In Fig. \ref{thermalization}(a), we show the projection of the wave function for $\mathcal{A}$-$\mathcal{R}$ system into the four Bell states. One can see that as the system evolves, the populations on the four Bell states gradually approach $1/4$ equally. When all four populations become $1/4$, it indicates that the reduced density matrix for the spin $\mathcal{A}$-$\mathcal{R}$ subsystem is identity. That is to say, all the initial state information about spins is lost. Precisely speaking, the initial spin information is scrambled into the cavity field, which has a much larger Hilbert space dimension. 

In Fig. \ref{thermalization}(b), we calculate the OTOC between $\sigma_x$ and $\sigma_z$ for the same initial state. The time scale that OTOC decays to nearly zero defines the information scrambling time in a quantum many-body system. Strictly speaking, the information scrambling time is usually longer than local thermalization time. However, since the Hilbert space for a single spin is quite small, they happen in nearly the same time scale. If one considers the OTOC between spin and cavity field, the scrambling time is longer because it takes more time for spin information to be scrambled into the large Hilbert space of cavity field.   

Another notable feature in Fig. \ref{thermalization}(b) is how the scrambling time depends on $g$. We show that the scrambling time for an intermediate $g$ is shorter than the scrambling times for both large $g$ and small $g$. This is because the system becomes close to integrable both in small and in large $g$ limits. At the moderate $g$ the competition effect between the coupling term and other two terms is the most significant, which renders the system the most chaotic one in this regime. Similar results have been obtained for the Bose-Hubbard model \cite{Shen} and for spin-model in a transverse field \cite{Spin}. In the view point of holographic duality, the regimes around the quantum critical point is most likely to have a holographic dual to a black hole, and therefore possess a larger Layponov exponent and a smaller scrambling time \cite{Shen}.   

\begin{figure}[t]
\centering
\includegraphics[width=0.95\linewidth]{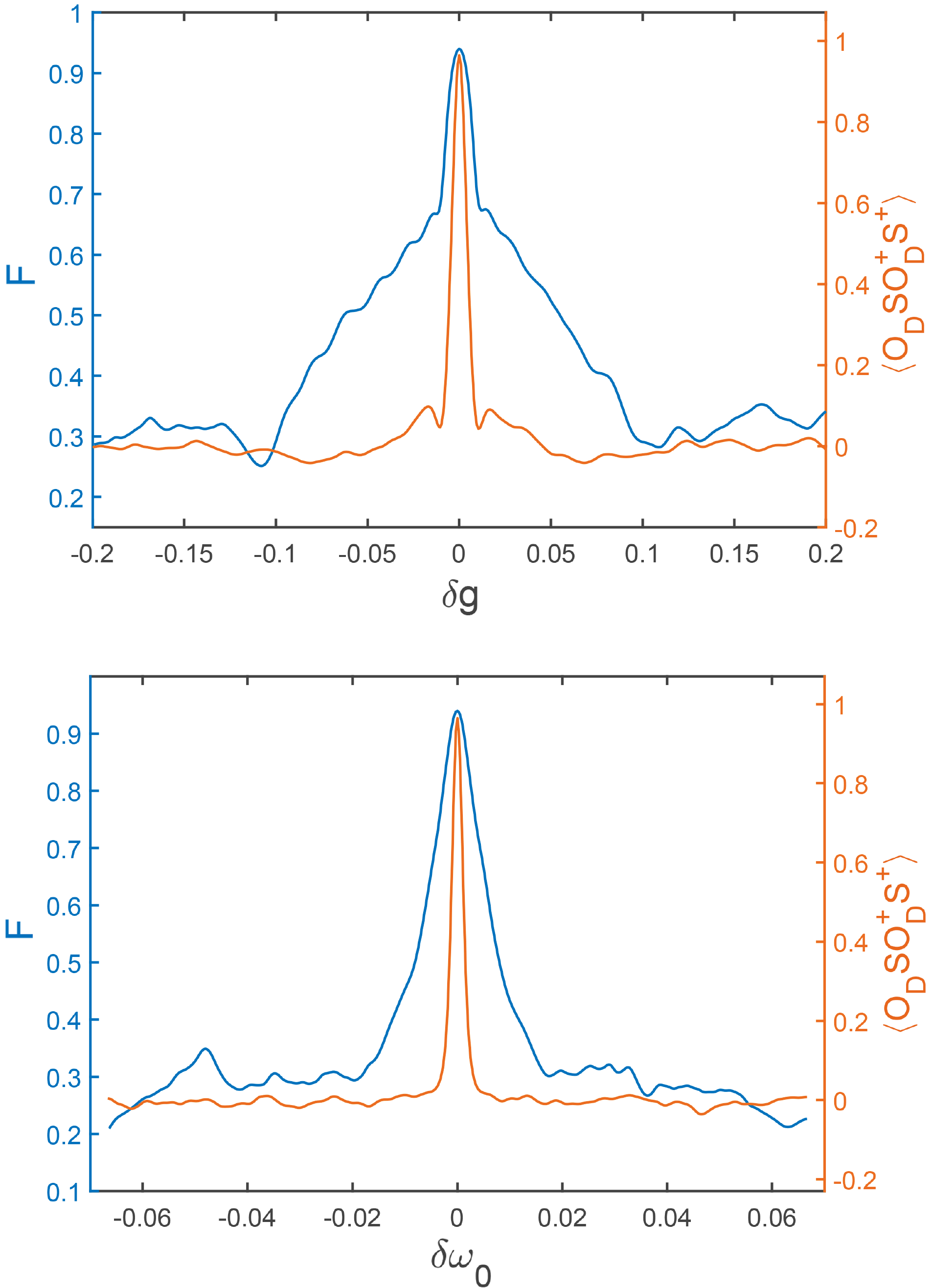}
\caption{The conditional probability $\mathcal{F}$ for $\delta g$ (a) and $\delta\omega_0$ (b), where $\delta g$ and $\delta \omega_0$ is the difference in coupling constant and difference in cavity photon frequency, respectively. We also plot the correlation function $\langle \hat{\mathcal{O}}_\mathcal{D} \hat{\mathcal{S}}\hat{\mathcal{O}}^\dag_\mathcal{D} \hat{\mathcal{S}}^\dag  \rangle$ that is closely related this dependence. }
\label{stability}
\end{figure}

\textit{Efficiency of the YKHP Protocol.} As discussed above, the key of the YKHP protocol is that when $\mathcal{D}$ and $\mathcal{D}^\prime$ is projected into an EPR state by measurement, $\mathcal{R}$ and $\mathcal{R}^\prime$ can recover the EPR correlation initially encoded between $\mathcal{A}$ and $\mathcal{R}$. Here we will investigate that in our practical model, how this conditional probability depends on time and coupling strength. Here we need to first properly define the Hilbert space $\mathcal{D}$. 

Considering a Fock state $|n\rangle_\text{L}$ in the left cavity, we can formally write $|n\rangle_\text{L}=|d_\mathcal{D}\times n_\mathcal{C}+n_\mathcal{D}\rangle_\text{L}  \equiv |n_\mathcal{C},n_\mathcal{D}\rangle_\text{L}$. Thus, we can define a $d_\mathcal{D}$-dimensional Hilbert space $\mathcal{D}$ as $\{|n_\mathcal{D}\rangle, n_\mathcal{D}=0,\dots, d_\mathcal{D}-1\}$. Here we should take $d_\mathcal{D}\hbar\omega\ll k_\text{B}T$ to ensure that $d_\mathcal{D}$ is much smaller than the dimension of occupied Hilbert space of $\mathcal{B}$. For example, take $d_\mathcal{D}=8$ when $k_\text{B}T=100$, we can identify the Hilbert space $\mathcal{D}$ as a product of three pseudo-spins-$1/2$, denoted by $\tau^1_\text{L}$, $\tau^2_\text{L}$ and $\tau^3_\text{L}$, and we can take $|0\rangle=|\uparrow\rangle_{\tau^1_\text{L}}|\uparrow\rangle_{\tau^2_\text{L}}|\uparrow\rangle_{\tau^3_\text{L}}$, $|1\rangle=|\downarrow\rangle_{\tau^1_\text{L}}|\uparrow\rangle_{\tau^2_\text{L}}|\uparrow\rangle_{\tau^3_\text{L}}$, ..., $|7\rangle=|\downarrow\rangle_{\tau^1_\text{L}}|\downarrow\rangle_{\tau^2_\text{L}}|\downarrow\rangle_{\tau^3_\text{L}}$. Similarly, we can define $\mathcal{D}^\prime$ as a part of the right system, and introduce three pseudo-spin $\tau^1_\text{R}$, $\tau^2_\text{R}$ and $\tau^3_\text{R}$ for $\mathcal{D}^\prime$.  Therefore, the EPR state between  $\mathcal{D}$ and $\mathcal{D}^\prime$ is defined as \cite{supple}
\begin{equation}
|\Psi^\text{EPR}_{\mathcal{D}\mathcal{D^\prime}}\rangle=\prod\limits_{i=1,2,3}\left(|\uparrow\rangle_{\tau^i_\text{L}}|\uparrow\rangle_{\tau^i_\text{R}}+|\downarrow\rangle_{\tau^i_\text{L}}|\downarrow\rangle_{\tau^i_\text{R}}\right).
\end{equation}

Our physical simulation of YKHP protocol works as follows: (i) we start with the initial state as
\begin{equation}
|\Phi(t=0)\rangle=|\Psi^\text{EPR}_{\mathcal{A}\mathcal{R}}\rangle\otimes |\Psi^\text{EPR}_{\mathcal{A}^\prime\mathcal{R}^\prime}\rangle\otimes |\text{TFD}_{\mathcal{B}\mathcal{B}^\prime}\rangle;
\end{equation}
and (ii) we evolve the state with two Dicke Hamiltonians and reach $|\Phi(t)\rangle$. (iii) We project $|\Phi(t)\rangle$ onto the EPR state between  $\mathcal{D}$ and $\mathcal{D}^\prime$ and the wave function on the remaining Hilbert space reads
\begin{equation}
|\tilde{\Phi}(t)\rangle=\langle{}\Psi^\text{EPR}_{\mathcal{D}\mathcal{D^\prime}}|\Phi(t)\rangle,
\end{equation}
and we can further project the state onto the EPR state between  $\mathcal{R}$ and $\mathcal{R}^\prime$ and the wave function of the rest $\mathcal{C}$ and $\mathcal{C'}$ Hilbert space reads
\begin{equation}
|\tilde{\tilde{\Phi}}(t)\rangle=\langle\Psi{}^\text{EPR}_{\mathcal{R}\mathcal{R^\prime}}|\tilde{\Phi}(t)\rangle.
\end{equation}
(iv) We compute the conditional probability $\mathcal{F}$ as 
\begin{equation}
\mathcal{F}=\frac{\langle\tilde{\tilde{\Phi}}(t) |\tilde{\tilde{\Phi}}(t)\rangle}{ \langle\tilde{\Phi}(t)|\tilde{\Phi}(t)\rangle}.
\end{equation}

The results of $\mathcal{F}$ is shown in Fig. \ref{conditional_probability}(a), with the emphasis on how $\mathcal{F}$ depends on $g$. First, $\mathcal{F}$ increases with the increasing of time $t$, and it saturates at long time. In Fig. \ref{conditional_probability}(b) we compare the saturation time of $\mathcal{F}$ with the scrambling time of an OTOC between spin operator and an operator $\hat{\mathcal{O}}_\mathcal{D}$ in the Hilbert space $\mathcal{D}$, defined as $\tau^{1,x}_\text{L} \otimes \tau^{2,y}_\text{L}\otimes \tau^{3,z}_\text{L} $, and find that these two time scales are consistent \cite{YK}. Thus, as discussed above, the scrambling time is shorter for an intermediate $g$, and concequently, $\mathcal{F}$ also saturates in a relatively shorter time scale. Secondly, the long time saturation value of $\mathcal{F}$ also depends on $g$ non-monotonically, as shown in the inset of Fig. \ref{conditional_probability}(a). It reaches a maximum value of $\sim 0.93$ for $g$ is around $2\sim 3$, where the system is in the most chaotic regime. That is the regime where the quantum evolution most closely resembles a Haar random unitary evolution.  

\textit{Stability of the YKHP Protocol.} Now we consider the situation that the parameters in $\hat{H}_\text{L}$ is not exactly the opposite of $\hat{H}_\text{R}$, which was previously considered in \cite{Yao1,Yao2}. Aside from the minus sign, either their coupling constant $g$ differs by $\delta g$ or the photon energy differs by $\delta\omega_0$. In Fig. \ref{stability} we show how the long time saturation value of $\mathcal{F}$ depends on either $\delta g$ or $\delta \omega_0$. It shows that on one hand, $\mathcal{F}$ decreases from being close to unity to about $1/4$ with the increasing of the absolute value of either $\delta g$ or $\delta \omega_0$. $1/4$ means that the EPR correlation between $\mathcal{R}$ and $\mathcal{R}^\prime$ is completely uncorrelated with the EPR correlation between $\mathcal{D}$ and $\mathcal{D}^\prime$. When the system is projected into the EPR state of $\mathcal{D}$ and $\mathcal{D}^\prime$, the system has equal probability to populate four Bell state of $\mathcal{R}$ and $\mathcal{R}^\prime$. 

To quantify how fast the conditional probability $\mathcal{F}$ decays with the difference between two system's Hamiltionian increasing, we prove an identity that \cite{supple}
\begin{equation}
\mathcal{F}=\frac{ \sum_{\hat{\mathcal{O}}_\mathcal{D}} \langle \hat{\mathcal{O}}_\mathcal{D} \hat{\mathcal{S}}\hat{\mathcal{O}}^\dag_\mathcal{D} \hat{\mathcal{S}}^\dag  \rangle}{d^2_\mathcal{A}-1+\sum_{\hat{\mathcal{O}}_\mathcal{D}} \langle \hat{\mathcal{O}}_\mathcal{D} \hat{\mathcal{S}}\hat{\mathcal{O}}^\dag_\mathcal{D} \hat{\mathcal{S}}^\dag  \rangle}, \label{F_stability}
\end{equation}
 where $d_\mathcal{A}$ is the Hilbert space dimension of system-$\mathcal{A}$ and the summation over $\hat{\mathcal{O}}_\mathcal{D}$ runs over a complete set of operators in the Hilbert space $\mathcal{D}$. Here $\hat{\mathcal{S}}$ denotes $\hat{U}_\text{L}\hat{U}^\dag_\text{R}$, where $\hat{U}_\text{L}$ denotes the evolution under $\hat{H}_\text{L}$ and $\hat{U}^*_\text{R}$ denotes the evolution under $\hat{H}_\text{R}$. It is clear that, if $\hat{H}_\text{R}$ and $\hat{H}_\text{L}$ are opposite with each other, then $\hat{U}_\text{L}=\hat{U}_\text{R}$ and $\hat{\mathcal{S}}$ is identity. Then $\sum_{\hat{\mathcal{O}}_\mathcal{D}} \langle \hat{\mathcal{O}}_\mathcal{D} \hat{\mathcal{O}}^\dag_\mathcal{D}  \rangle =d^2_\mathcal{D}$. Since $d_\mathcal{D}\gg d_\mathcal{A}$, we have $\mathcal{F}\rightarrow 1$. If $\hat{H}_\text{R}$ and $\hat{H}_\text{L}$ become two very different Hamiltonians, then $\hat{\mathcal{S}}\hat{\mathcal{O}}^\dag_\mathcal{D} \hat{\mathcal{S}}^\dag$ and $\hat{\mathcal{O}}_\mathcal{D}$ are two different operators, and $\langle \hat{\mathcal{O}}_\mathcal{D} \hat{\mathcal{S}}\hat{\mathcal{O}}^\dag_\mathcal{D} \hat{\mathcal{S}}^\dag  \rangle$ vanishes except for $\hat{\mathcal{O}}_\mathcal{D}$ being identity. Then $\mathcal{F}$ approaches $1/d^2_\mathcal{A}$. In this case, since $\mathcal{A}$ system is a single spin-$1/2$ and $d_\mathcal{A}=2$. That also explains why $\mathcal{F}$ approaches $1/4$ when parameters of two Hamiltonians become different. 
 
\textit{Conclusion.} In summary, we have presented a physical realization of the Hayden-Preskill protocol in two copies of Dicke models with the help of the thermofield double state. It is interesting to note the dual role of information scrambling regarding decoding the initial state information. In a single system, it is the information scrambling that prevents decoding the initial state information from local measurements after thermalization. But in the thermofield double system, it is also the information scrambling that allows the YKHP protocol to read out the initial state information. We show that in our model, the decoding efficiency reaches a maximum close to unity for intermediate spin-cavity coupling $g$, where the system is most chaotic and the scrambling is the fastest. We hope that our results can simulate experimental simulation of the HP thought experiment in synthetic quantum systems.   

\textit{Acknowledgment.}   This work is supported by Beijing Distinguished Young Scientist Program (HZ), MOST under Grant No. 2016YFA0301600 (HZ) and NSFC Grant No. 11734010 (HZ and YC), No. 11604225 (YC) and Beijing Natural Science Foundation (Z180013) (YC). PZ acknowledges support from the Walter Burke Institute for Theoretical Physics at Caltech.

\end{document}